# HoloClean: Holistic Data Repairs with Probabilistic Inference

Theodoros Rekatsinas*, Xu Chu†, Ihab F. Ilyas†, Christopher Ré*
* Stanford University and † University of Waterloo

## ABSTRACT

We introduce HoloClean, a framework for holistic data repairing driven by probabilistic inference. HoloClean unifies existing qualitative data repairing approaches, which rely on integrity constraints or external data sources, with quantitative data repairing methods, which leverage statistical properties of the input data. Given an inconsistent dataset as input, HoloClean automatically generates a probabilistic program that performs data repairing. Inspired by recent theoretical advances in probabilistic inference, we introduce a series of optimizations which ensure that inference over HoloClean's probabilistic model scales to instances with millions of tuples. We show that HoloClean scales to instances with millions of tuples and find data repairs with an average precision of $\sim 90\%$ and an average recall of above $\sim 76\%$ across a diverse array of datasets exhibiting different types of errors. This yields an average F1 improvement of more than $2\times$ against state-of-the-art methods.

## 1. INTRODUCTION

The process of ensuring that data adheres to desirable quality and integrity constraints (ICs), referred to as *data cleaning*, is a major challenge in most data-driven applications. Given the variety and voluminous information involved in modern analytics, the need for identifying and repairing inconsistencies caused by incorrect, missing, and duplicate data, has become even more pronounced. As a result, large-scale data cleaning has re-emerged as the key goal of many academic [8, 22, 24] and industrial efforts (including Tamr [38], Trifacta Wrangler [25], and many more).

Data cleaning can be separated in two tasks: (i) *error detection*, where data inconsistencies such as duplicate data, integrity constraint violations, and incorrect or missing data values are identified, and (ii) *data repairing*, which involves updating the available data to remove any detected errors. Significant efforts have been made to automate both tasks, and several surveys summarize these results [18, 24, 33]. For error detection, many methods rely on violations of integrity constraints [8, 11] or duplicate [20, 29, 32] and outlier detection [15, 22] methods to identify errors. For data repairing, state-of-the-art methods use a variety of signals: (i) integrity constraints [6, 12], (ii) external information [13, 19], such as dictionaries, knowledge bases, and annotations by human experts, or (iii) statistical profiling of the input dataset [31, 39].

While, ensembles of automatic error detection methods were shown to achieve precision and recall greater than 0.6 and 0.8 for multiple real-world datasets [2], this is not the case with automatic data repairing [23]. We evaluated state-of-the-art repairing methods [12, 13, 39] on different real-world datasets (Section 6) and found that (i) their average F1-score (i.e., the harmonic mean of precision and recall) across datasets is below 0.35, and (ii) in many cases these methods did not perform any correct repairs. This is because these methods limit themselves to only one of the aforementioned signals, and ignore additional information that is useful to identify the correct value of erroneous records. In this paper, we show that if we combine these signals in a unified framework, we obtain data repairs with an average F1-score of more than 0.8. We use a real-world dataset to demonstrate the limitations of existing data repairing methods and motivate our approach.

**Example 1.** *We consider a dataset from the City of Chicago[1] with information on inspections of food establishments. A snippet is shown in Figure 1(A). The dataset is populated by transcribing forms filled out by city inspectors, and as such, contains multiple errors. Records can contain misspelled entries, report contradicting zip codes, and use different names for the same establishment.*

Figure 1 shows instances of the aforementioned signals. In our example we have access to a set of functional dependencies (see Figure 1(B)) and an external dictionary of address listings in Chicago (Figure 1(D)). Co-occurrence statistics can also be obtained by analyzing the original input dataset in Figure 1(A).

First, we focus on data repairing methods that rely on integrity constraints [5, 8, 12]. These methods assume the majority of input data to be clean and use the principle of *minimality* [3, 10, 17] as an operational principle to perform repairs. The goal is to update the input dataset such that no integrity constraints are violated. Informally, minimality states that given two candidate sets of repairs, the one with fewer changes with respect to the original data is preferable. Nevertheless, minimal repairs do not necessarily correspond to correct repairs: An example minimal repair is shown in Figure 1(E). This repair chooses to update the zip code of tuple $t1$ so that all functional dependencies in Figure 1(B) are satisfied. This particular repair introduces an error as the updated zip code is wrong. This approach also fails to repair the zip code of tuples $t2$ and $t3$ as well as the "DBAName" and "City" fields of tuple $t4$ since altering those leads to a non-minimal repair.

Second, methods that rely on external data [13, 19] match records of the original dataset to records in the external dictionaries or knowledge bases to detect and repair errors in the former. The matching process is usually described via a collection of *matching dependencies* (see Figure 1(C)) between the original dataset and external information. A repair using such methods is shown in Figure 1(F). This repair fixes most errors but fails to repair the "DBAName" field of tuple $t4$ as no information for this field is provided in the external data. In general, the quality of repairs performed by methods that use external data can be poor due to the limited coverage of external resources or these methods may not be applicable as for many domains a knowledge base may not exist.

---
[1] https://data.cityofchicago.org

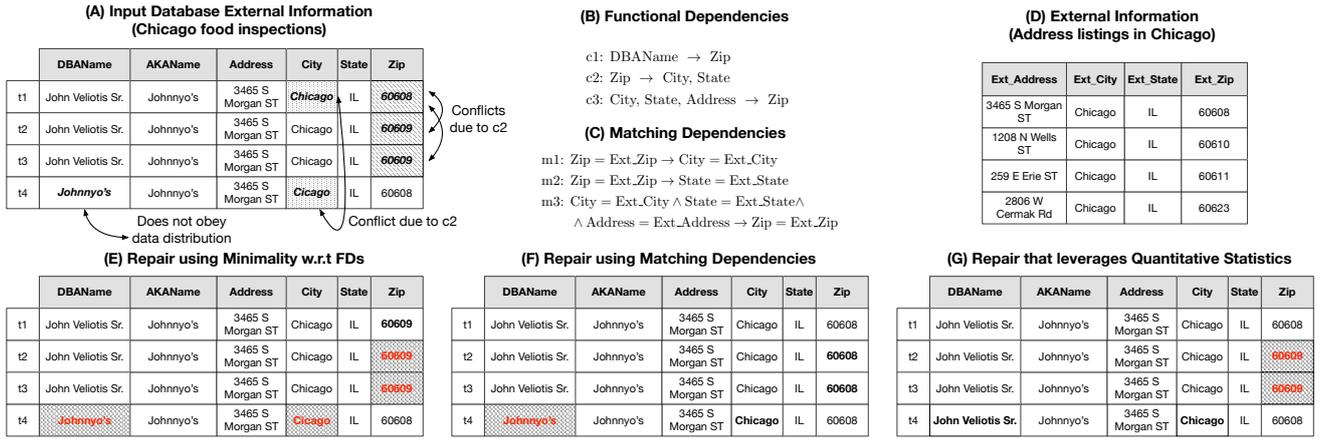

Figure 1: A variety of signals can be useful for data cleaning: integrity constraints, external information in the form of dictionaries, and basic quantitative statistics of the dataset to be cleaned. Using each signal in isolation can lead to repairs that do not fix all errors in the input data or even introduce new errors.

Finally, data repairing methods that are based on statistical analysis [31, 39], leverage quantitative statistics of the input dataset, e.g., co-occurrences of attribute values and the empirical distribution characterizing attributes of the input dataset, and use those to clean the input dataset. These techniques overlook integrity constraints. Figure 1(G) shows such a repair. As shown the "DBAName" and "City" fields of tuple $t4$ are updated as their original values correspond to outliers with respect to other tuples in the dataset. However, this repair does not have sufficient information to fix the zip code of tuples $t2$ and $t3$.

In our example, if we combine repairs that are based on different signals, we can repair all errors in the input dataset correctly. If we combine the zip code and city repairs from Figure 1(F) with the DBAName repair from Figure 1(G) we can repair all inaccuracies in the input dataset. Nonetheless, combining heterogeneous signals can be challenging. This is not only because each type of signal is associated with different operations over the input data (e.g., integrity constraints require reasoning about the satisfiability of constraints while external information requires efficient matching procedures) but different signals may suggest conflicting repairs. For instance, if we naively combine the repairs in Figure 1 we end up with conflicts on the zip code of tuples $t2$ and $t3$. The repairs in Figure 1(E) and (G) assign value "60609" while the repair in Figure 1(F) assigns value "60608". This raises the main question we answer in this paper: *How can we combine all aforementioned signals in a single unified data cleaning framework, and which signals are useful for repairing different records in an input dataset?*

*Our Approach.* We introduce HoloClean, the first data cleaning system that unifies integrity constraints, external data, and quantitative statistics, to repair errors in structured data sets (see Table 1). Instead of relying on each signal solely to perform data repairing, we use all available signals to suggest data repairs. We consider the input dataset as a *noisy version* of a hidden clean dataset and treat each signal as *evidence* on the correctness of different records in that dataset. To combine different signals, we rely on probability theory as it allows us to reason about inconsistencies across those.

Table 1: HoloClean compared to other data cleaning methods.

| System | Integrity Constraints | External Data | Statistical Profiles |
|---|---|---|---|
| Holistic [12] | ✓ | - | - |
| KATARA [13] | - | ✓ | - |
| SCARE [39] | - | - | ✓ |
| HoloClean | ✓ | ✓ | ✓ |

HoloClean automatically generates a probabilistic model [27] whose random variables capture the uncertainty over records in the input dataset. Signals are converted to features of the graphical model and used to describe the distribution characterizing the input dataset. To repair errors, HoloClean uses statistical learning and probabilistic inference over the generated model.

HoloClean exhibits significant improvements over state-of-the-art data cleaning methods: we show that across multiple datasets HoloClean finds repairs with an average precision of $\sim 90\%$ and an average recall of $\sim 76\%$, obtaining an average F1-score improvement of more than $2\times$ against state-of-the-art data repairing methods. Specifically, we find that combining all signals yields an F1-score improvement of $2.7\times$ against methods that only use integrity constraints, an improvement of $2.81\times$ against methods that only leverage external information, and an improvement of $2.29\times$ against methods that only use quantitative statistics.

**Technical Challenges.** Though probabilistic models provide a means for unifying all signals, it is unclear that probabilistic inference scales to large, complex data repairing instances. A probabilistic inference program involves two tasks: (i) *grounding*, which enumerates all possible interactions between correlated random variables to materialize a *factor graph* that represents the joint distribution over all variables, and (ii) *inference* where the goal is to compute the marginal probability for every random variable. These tasks are standard but non-trivial:

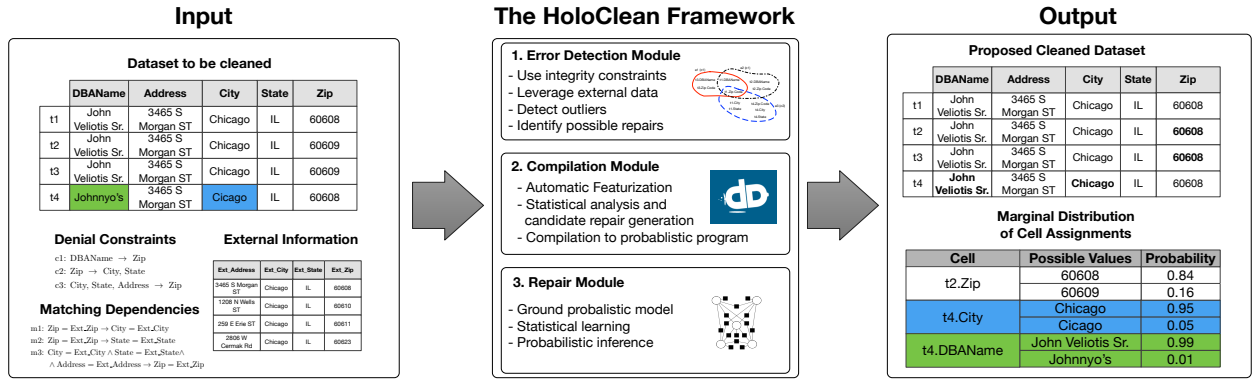

Figure 2: An overview of HoloClean. The provided dataset along with a set of denial constraints is compiled into a declarative program which generates the probabilistic model used to solve data repairing via statistical learning and probabilistic inference.

(1) Integrity constraints that span multiple attributes can cause combinatorial explosion problems. Grounding the interactions due to integrity constraints requires considering all value combinations that attributes of erroneous tuples can take. If attributes are allowed to obtain values from large domains, inference can become intractable. For example, we consider repairing the smallest dataset in our experiments, which contains 1,000 tuples, and allow attributes in erroneous tuples to obtain any value from the set of consistent assignments present in the dataset. Inference over the resulting probabilistic model does not terminate after an entire day. Thus, we need mechanisms that limit the possible value assignments for records that need to be repaired by HoloClean's probabilistic model.

(2) Integrity constraints introduce correlations between pairs of random variables associated with tuples in the input dataset. Enumerating these interactions during grounding results in factor graphs of quadratic size in the number of tuples. For example, in our experiments we consider a dataset with more than two million tuples. Enforcing the integrity constraints over all pairs of tuples, yields a factor graph with more than four trillion interactions across random variables. It is obvious that grounding this factor graph requires an unrealistic amount of time. Thus, we need to avoid evaluating integrity constraints for pairs of tuples that cannot result in violations.

(3) Finally, probabilistic inference is #P-complete in the presence of complex correlations, such as hard constraints. Thus, approximate inference techniques such as Gibbs sampling are required. In the presence of complex correlations, Gibbs sampling is known to require an exponential number of samples in the number of random variables to *mix* [36], i.e., reach a stationary distribution. Nevertheless, recent theoretical advancements [9, 36] in statistical learning and inference show that relaxing hard constraints to soft constraints introduces a trade-off between the computational efficiency and the quality of solutions obtained. This raises the technical question, how to soften hard integrity constraints to obtain scalable probabilistic models that still obtain accurate results for data repairing.

**Technical Contributions.** Our main technical contributions are:

(1) We design a compiler that automatically generates a probabilistic model for repairing a dataset. The output model defines a factor graph that unifies different signals for data repairing, including integrity constraints, external data, and quantitative statistics of the input dataset. Our compiler supports different error detection methods, such as constraint violation detection and outlier detection.

(2) We design an algorithm that uses Bayesian analysis to prune the domain of the random variables corresponding to noisy cells in the input dataset. This algorithm allows us to systematically trade-off the scalability of HoloClean and the quality of repairs obtained by it. We also introduce a scheme that partitions the input dataset into non-overlapping groups of tuples and enumerates the correlations introduced by integrity constraints only for tuples within the same group. Using this scheme, HoloClean avoids grounding factors graphs of quadratic size. Empirically, we find that these two optimizations reduce the size of factor graphs generated by HoloClean between $7\times$ (for small datasets) and $96,000\times$ (for the largest dataset considered), and allow HoloClean to scale to inputs with millions of tuples.

(3) We introduce an approximation scheme that relaxes hard integrity constraints to priors over the random variables in Holo-Clean's probabilistic model. This relaxation results in a probabilistic model with *independent random variables* for which Gibbs sampling only requires a polynomial number of samples to mix. We empirically study the trade-off between the runtime and quality of repairs obtained when relaxing integrity constraints. We show that our approximation not only leads to more scalable data repairing models but also results in repairs of the same quality as those obtain by non-approximate models.

*Outline.* In Section 2 we formalize the problem that HoloClean addresses and provide an overview of HoloClean. In Section 3 we review necessary background material. In Sections 4 and 5, we introduce the main compilation routine of HoloClean and optimizations for scaling probabilistic inference in HoloClean. Finally, in Section 6 we present an experimental evaluation of HoloClean and competing data repairing methods, and conclude in Section 7.

## 2. THE HoloClean FRAMEWORK

We formalize the goal of HoloClean and provide an overview of HoloClean's solution to data repairing.

### 2.1 Problem Statement

The goal of HoloClean is to identify and repair erroneous records in a structured dataset $D$. We denote $A = \{A_1, A_2, \ldots, A_N\}$ the attributes that characterize dataset $D$. We represent $D$ as a set of tuples, where each tuple $t \in D$ is a set of cells denoted as $Cells[t] = \{A_i[t]\}$. Each cell $c$ corresponds to a different attribute in $A$. For example, the dataset in Figure 1 has attributes, "DBAName", "AKAName", "Address", "City", "State", and "Zip", and consists of four tuples. We denote $t[A_n]$ the $n$-th cell of tuple $t$ for attribute $A_n \in A$.

We assume that errors in $D$ occur due to inaccurate cell assignments, and we seek to repair errors by updating the values of cells in $D$. This is a typical assumption in many data cleaning systems [12, 13, 31, 40]. For each cell we denote by $v_c^*$ its unknown true value and by $v_c$ its initial observed value. We use $\Omega$ to denote the initial observed values for all cells in $D$. We term an *error* in $D$ to be each cell $c$ with $v_c \neq v_c^*$. The goal of HoloClean is to estimate the latent true values $v_c^*$ for all erroneous cells in $D$. We use $\hat{v}_c$ to denote the estimated true value of a cell $c \in D$. We say that an inaccurate cell is correctly repaired when $\hat{v}_c = v_c^*$.

## 2.2 Solution Overview

An overview of HoloClean is shown in Figure 2. HoloClean takes as input a dirty database instance $D$, along with a set of available repairing constraints $\Sigma$. In the current implementation we limit these constraints to: (i) *denial constraints* [10] that specify various business logic and integrity constraints (we review denial constraints in Section 3.1), and (ii) *matching dependencies* [5, 13, 19] to specify lookups to available external dictionaries or labeled (clean) data. We briefly review denial constraints in Section 3.1. Given the aforementioned input, HoloClean's workflow follows three steps:

**Error Detection.** The first step in the workflow of HoloClean is to detect cells in $D$ with potentially inaccurate values. This process separates $D$ into *noisy* and *clean* cells, denoted $D_n$ and $D_c$, respectively. HoloClean treats error detection as a black box. Users have the flexibility to use any method that detects erroneous cells. The output of such methods is used to form $D_n$ and $D_c$ is set to $D_c = D \setminus D_n$. Our current implementation included a series of error detection methods, such as methods that leverage denial constraints to detect erroneous cells [11], outlier detection mechanisms [15, 22], and methods that rely on external and labeled data [5, 13, 19].

**Compilation.** Given the initial observed cell values $\Omega$ and the set of repairing constraints $\Sigma$, HoloClean follows probabilistic semantics to express the uncertainty over the value of noisy cells. Specifically, it associates each cell $c \in D$ with a random variable $T_c$ that takes values form a finite domain $dom(c)$, and compiles a probabilistic graphical model that describes the distribution of random variables $T_c$ for cells in $D$. HoloClean relies on factor graphs [27] to represent the probability distribution over variables $T_c$. HoloClean is built on top of DeepDive [37], a declarative probabilistic inference engine. In Section 3.2, we review factor graphs and how probabilistic models are defined in DeepDive.

**Data Repairing.** To repair $D$, HoloClean runs statistical learning and inference over the joint distribution of variables $T_1, T_2, \ldots$ to compute the marginal probability $P(T_c = d; \Omega, \Sigma)$ for all values $d \in dom(c)$, and assigns $\hat{v}_c$ to the value that maximizes the probability of variable $T_c$. Let $T$ be the set of all variables $T_c$. HoloClean uses empirical risk minimization (ERM) over the likelihood $\log P(T)$ to compute the parameters of its probabilistic model. Variables that correspond to clean cells in $D_c$ are treated as evidence and are used to learn the parameters of the model. Variables for noisy cells in $D_n$ correspond to query variables whose value needs to be inferred. Efficient methods such as stochastic gradient descent (SGD) are used to optimize over that objective. Approximate inference via Gibbs sampling [41] is used to estimate the value $\hat{v}_c$ of noisy cells. Variables $\hat{v}_c$ are assigned to the maximum a posteriori (MAP) estimates of variables $T_c$.

Similar to existing automatic data repairing approaches, HoloClean's recall is limited by the error detection methods used. Error detection is out of the scope of this paper. However, as part of future directions (Section 7), we discuss how state-of-the-art weak supervision methods can be used to improve error detection.

Finally, we point out that each repair proposed by HoloClean is associated with a marginal probability that carries rigorous semantics. For example, if the proposed repair for a record in the initial dataset has a probability of 0.6 it means that HoloClean is 60% confident about this repair. Intuitively, this means that if HoloClean proposes 100 repairs then only 60 of them will be correct. As a result, we can use these marginal probabilities to solicit user feedback. For example, we can ask users to verify repairs with low marginal probabilities and use those as labeled examples to retrain the parameters of HoloClean's model using standard incremental learning and inference techniques [37].

## 3. BACKGROUND

We review concepts and terminology used in the next sections.

### 3.1 Denial Constraints

HoloClean allows users to specify a set of *integrity constraints* to ensure the consistency of data entries in $D$. In our current implementation these constraints correspond to the family of denial constraints [10]. Denial constraints subsume several types of integrity constraints such as functional dependencies, conditional functional dependencies [8], and metric functional dependencies [28].

Given a set of operators $B = \{=, <, >, \neq, \leq, \approx\}$, with $\approx$ denoting similarity, denial constraints are first-order formulas over *cells* of tuples in dataset $D$. Denial constraints take the form $\sigma : \forall t_i, t_j \in D : \neg(P_1 \land \cdots \land P_k \land \cdots \land P_K)$ where each predicate $P_k$ is of the form $(t_i[A_n] \circ t_j[A_m])$ or $(t_i[A_n] \circ \alpha)$ where $A_n, A_m \in A$, $\alpha$ denotes a constant and $\circ \in B$. We illustrate this with an example:

**Example 2.** *Consider the functional dependency Zip $\rightarrow$ City, State from the food inspection dataset in Figure 1. This dependency can be represented using the following two denial constraints:*

$$\forall t_1, t_2 \in D : \neg(t1[Zip] = t2[Zip] \land t1[City] \neq t2[City])$$
$$\forall t_1, t_2 \in D : \neg(t1[Zip] = t2[Zip] \land t1[State] \neq t2[State])$$

### 3.2 Factor Graphs

A factor graph is a hypergraph $(T, F, \theta)$ in which $T$ is a set of nodes that correspond to random variables and $F$ is a set of hyperedges. Each hyperedge $\phi \in F$, where $\phi \subseteq T$, is referred to as a *factor*. For ease of exposition only, we assume that all variables $T$ have a common domain $\mathbb{D}$. Each hyperedge $\phi$ is associated with a *factor function* a real-valued weight $\theta_\phi$ and takes an assignment of the random variables in $\phi$ and return a value in $\{-1, 1\}$ (i.e., $h_\phi : \mathbb{D}^{|f|} \rightarrow \{-1, 1\}$). Hyperedges $f$, functions $h_\phi$, and weights $\theta_\phi$ define a *factorization* of the probability distribution $P(T)$ as:

$$P(T) = \frac{1}{Z} \exp\left(\sum_{\phi \in F} \theta_\phi \cdot h_\phi(\phi)\right) \qquad (1)$$

where $Z$ is called the *partition function* and corresponds to a constant ensuring we have a valid distribution.

Recently, declarative probabilistic frameworks, such as DeepDive [37], Alchemy [1], and PSL [4] have been introduced to facilitate the construction of large scale factor graphs. In HoloClean, we choose to use DeepDive. In DeepDive, users can specify factor graphs via *inference rules* in DDlog, a custom declarative language that is semantically similar to Datalog but extends it to encode probability distributions. A probabilistic model in DeepDive corresponds to a collection of rules in DDlog.

DDlog rules are specified over relations in an input database using Datalog-like statements. For example, the following DDlog rule states that the tuples of relation $Q$ are derived from $R$ and $S$, where the second column of $R$ is unified with the first column of $S$, i.e., the body is a equi-join between $R$ and $S$.

$$Q(x, y) : -R(x, y), S(y), [x = \text{``a''}]$$

Here, $Q(x, y)$ is the *head* of the rule, and $R(x, y)$ and $S(y)$ are *body atoms*. The body also contains a condition $[x = \text{``a''}]$ on the values that the first attribute of relation $R$ can take. Predicate $[x = \text{``a''}]$ is called the scope of the rule. Finally, $x$ and $y$ are variables of the rule. We next describe how such rules can be extended to define a factor graph.

Relations in DDlog can be augmented with a special question-mark annotation to specify random variables. For example, let Fact(x) be a relation containing facts that we want to infer if they are True or False. Let IsTrue?(x) be a relation such that each assignment to $x$ represents a different random variable taking the value True or False. The next DDlog rule defines this random variable relation:

$$\text{IsTrue?}(x) : -\text{Fact}(x)$$

Grounding relation Fact generates a random variable for each value of $x$. These correspond to nodes $T$ in the factor graph. In the remainder of the paper we refer to relations annotated with a question-mark as *random variable relations*.

Given relations that define random variables we can extend Datalog rules with weight annotations to encode *inference rules*, which express the factors of a factor graph. We continue with the previous example and let HasFeature(x, f) be a relation that contains information about the features $f$ that a fact $x$ can have. We consider the following inference rule:

$$\text{IsTrue?}(x) : -\text{HasFeature}(x, f) \text{ weight} = w(f)$$

The head of this rule defines a *factor function* that takes as input one random variable—corresponding to an assignment of variable $x$—and returns 1.0 when that variable is set to True and $-1.0$ otherwise. Effectively, this rule associates the features for each fact $x$ with its corresponding random variable. The weights are parameterized by variable $f$ to allow for different confidence levels across features. To generate the factors specified by the above rule we ground its body by evaluating the corresponding query. Grounding generates a factor (hyper-edge in the factor graph) for each assignment of variables $x$ and $f$. In general, the head of inference rules can be a complex boolean function that introduces correlations across random variables.

Finally, variables in the generated factor graph are separated in two types: a set $E$ of *evidence variables* (those fixed to a specific value) and a set $Q$ of *query variables* whose value needs to be inferred. During inference, the values of all weights $w$ are assumed to be known, while the goal of learning is to find the set of weights that maximizes the probability of the evidence.

## 4. COMPILATION IN HOLOCLEAN

HoloClean compiles all available signals, including denial constraints, external data, matching dependencies, and the minimality principle, to a DDlog program that defines the factor graph used to repair the input dataset $D$. The generated DDlog program contains: (i) rules that capture quantitative statistics of $D$; (ii) rules that encode matching dependencies over external data; (iii) rules that represent dependencies due to integrity constraints; (iv) rules that encode the principle of minimality. The groundings of these rules construct factors $h_\phi$ in Equation 1 as in described Section 3.2.

HoloClean's compilation involves two major steps: (i) first HoloClean generates relations used to form the body of DDlog rules, and then (ii) uses those relations to generate inference DDlog rules that define HoloClean's probabilistic model. We describe each of these steps in detail.

### 4.1 DDlog Relations in HoloClean

HoloClean generates several relations that correspond to transformations of the input dataset $D$. The following two variables are used to specify fields of these relations: (i) $t$ is a variable that ranges over the identifiers of tuples in $D$, and (ii) $a$ is a variable that ranges over the attributes of $D$. We also denote by $t[a]$ a cell in $D$ that corresponds to attribute $a$ of tuple $t$. HoloClean's compiler generates the following relations:

(1) Tuple(t) contains all identifiers of tuples in $D$.
(2) InitValue(t, a, v) maps every cell $t[a]$ to its initial value $v$.
(3) Domain(t, a, d) maps every cell $t[a]$ to the possible values it can take, where variable $d$ ranges over the domain of $t[a]$.
(4) HasFeature(t, a, f) associates every cell $t[a]$ with a series of features captured by variable $f$.

All relations are automatically populated. Relations Tuple, InitValue, and Domain are populated directly from the values in $D$, and the domain of each attribute in $D$. In Section 5.1.1, we show how to prune entries in Relation Domain for scalable inference. Finally, HasFeature is populated with two types of features: (i) given a cell $c$, HoloClean considers as features the values of other cells in the same tuple as $c$ (e.g., "Zip=60608"). These features capture distributional properties of $D$ that are manifested in the co-occurrences of attribute values; and (ii) if the provenance and lineage of $t[a]$ is provided (e.g., the source from which $t$ was obtained) we use this information as additional features. This allows HoloClean to reason about the trustworthiness of different sources [35] to obtain more accurate repairs. Finally, users have the flexibility to specify more features by adding tuples in Relation HasFeature.

To capture external data, HoloClean assumes an additional relation that is optionally provided as input by the user:

(5) ExtDict($t_k$, $a_k$, v, k) stores information from multiple external dictionaries identified by the indicator variable $k$. Similar to InitValue, variables $t_k$ and $a_k$ range over the tuples and attributes of dictionary $k$, respectively. Relation ExtDict maps each $t_k[a_k]$ to its value $v$ in Dictionary $k$.

### 4.2 Translating Signals to Inference Rules

HoloClean's compiler first generates a DDlog rule to specify the random variables associated with cells in the input dataset $D$:

$$\text{Value?}(t, a, d) : - \text{Domain}(t, a, d)$$

This rule defines a random variable relation Value?(t, a, d), which assigns a categorical random variable to each cell $t[a]$. Grounding this rule generates the random variables in HoloClean's probabilistic model. In the remainder of this section, we show how HoloClean expresses each repairing signal, described in Section 1, as an inference DDlog rule over these random variables. Grounding these rules populate the factors used in HoloClean's probabilistic model, which completes the specification of the full factor graph used for inferring the correct repairs.

*Quantitative Statistics.* We use the features of cells in $D$, stored in Relation HasFeature(t, a, f), to capture the quantitative statistics of dataset $D$. HoloClean generates the following inference rule to encode the effect of features on the assignment of random variables:

Value?(t, a, d) :− HasFeature(t, a, f) weight = w(d, f)

Weight $w(d, f)$ is parameterized by $d$ and $f$ to allow for different confidence levels per feature. These weights are learned using evidence variables that have fixed assignments (Section 2.2).

*External Data.* Given Relation ExtDict($t_k$, $a_k$, v, k), described in Section 4.1, along with a collection of matching dependencies—expressed as implications in first-order logic—HoloClean generates additional DDlog rules that capture the effect of the external dictionaries on the assignment of random variables. First, HoloClean generates DDlog rules to populate a relation Matched, which contains all identified matches. We use an example to demonstrate the form of DDlog rules used to populate Matched:

**Example 3.** *We consider the matching dependency between Zip and City from Example 1. HoloClean generates the DDlog rule:*

Matched(t1, City, c2, k) :−Domain(t1, City, c1), InitValue(t1, Zip, z1), ExtDict(t2, Ext_Zip, z1, k), ExtDict(t2, Ext_City, c2, k) , $[c1 \approx c2]$

*where $\approx$ is a similarity operator and $k$ is the indicator of the external dictionary used. The rule dictates that for a tuple $t1$ in $D$ if the zip code matches the zip code of a tuple $t2$ in the external dictionary $k$, then the city of $t1$ has to match the city of $t2$. The DDlog formula populates Matched with the tuple $\langle t1, City, c2 \rangle$, where $c2$ is the lookup value of $t1[City]$ in Dictionary $k$.*

HoloClean's compiler generates the following inference rule to capture the dependencies between the external dictionaries and the random variables using Relation Matched:

Value?(t, a, d) :− Matched(t, a, d, k) weight = w(k)

Weight $w(k)$ is parameterized by the identifier of the dictionary, $k$, to allow for different levels of reliability per dictionary.

---

**Algorithm 1:** Denial Constraint Compilation to DDlog Rules

**Input**: Denial constraints in $\Sigma$, constant weight $w$
**Output**: DDlog Rules for Denial Constraints
rules = [];
**for** *each constraint* $\sigma : \forall t_1, t_2 \in D : \neg(P_1 \wedge \cdots \wedge P_K)$ **do**
  /* Initialize the head and scope of the new DDlog rule*/
  $H \leftarrow \emptyset, S \leftarrow \emptyset$;
  **for** *each predicate $P_k$ in $\sigma$* **do**
    **if** $P_k$ *is of the form* $(t_1[A_n] \text{ o } t_2[A_m])$ **then**
      $H = H \cup \{\text{Value?}(t1, A_n, v_{1k}) \wedge \text{Value?}(t2, A_m, v_{2k})\}$;
      $S = S \cup \{v_{1k} \text{ o } v_{2k}\}$;
    **if** $P_k$ *is of the form* $(t_1[A_n] \text{ o } \alpha)$ **then**
      $H = H \cup \{\text{Value?}(t1, A_n, v_{1k})\}$;
      $S = S \cup \{v_{1k} \text{ o } \alpha\}$;
  rules += ! $\bigwedge_{h \in H} h$ :− Tuple(t1), Tuple(t2),[S] weight = w;
**return** rules;

---

*Dependencies From Denial Constraints.* As defined in Section 3.1, HoloClean accepts denial constraints of the form $\sigma : \forall t_1, t_2 \in D : \neg(P_1 \wedge \cdots \wedge P_K)$, where each predicate $P_k$ is of the form $(t_1[A_n] \text{ o } t_2[A_m])$ or $(t_1[A_n] \text{ o } \alpha)$, $A_n, A_m \in A$, $\alpha$ is a constant, and o is an operation. Denial constraints correspond to first-order logic formulas, thus, can be easily converted to DDlog using Algorithm 1. In Algorithm 1, the quantifier $\forall t_1, t_2$ of a denial constraint $\sigma$ is converted to a self-join Tuple(t1), Tuple(t2) over Relation Tuple in DDlog. We apply standard ordering strategies to avoid grounding duplicate factors. The details are omitted from the pseudocode for clarity. We illustrate the output of Algorithm 1 with an example:

**Example 4.** *Consider the denial constraint from Example 2:*

$$\forall t_1, t_2 \in D : \neg(t1[Zip] = t2[Zip] \wedge t1[State] \neq t2[State])$$

*This constraint can be expressed as a factor template in DDlog as:*

!(Value?(t1, Zip, z1) ∧ Value?(t2, Zip, z2)∧
Value?(t1, State, s1) ∧ Value?(t2, State, s2)) :−
Tuple(t1), Tuple(t2), $[z1 = z2, s1 \neq s2]$ $weight = w$

Setting $w = \infty$ converts these factors to hard constraints. However, probabilistic inference over a set of hard constraints is in general #-P complete, as it is related to model counting [16]. HoloClean allows users to relax hard constraints to *soft constraints* by assigning $w$ to a constant value. The larger the value of $w$ the more emphasis is put on satisfying the given denial constraints.

*Minimality Priors.* Using minimality as an operational principle might lead to inaccurate repairs [23]. However, minimality can be viewed as the prior that the input dataset $D$ contains fewer erroneous records than clean records. The stronger this prior is, the smaller the total number of updated cells will be. To capture the above prior, HoloClean generates the following DDlog rule:

Value?(t, a, d) :− InitValue(t, a, d) weight = w

Weight $w$ is a positive constant indicating the strength of this prior.

So far we showed how HoloClean maps various repairing signals into DDlog rules for constructing the full factor graph used for inference. The extensibility of HoloClean depends on our ability to map other external signals as additional DDlog rules. In the following section, we show how we manage the complexity of the generated factor graph to allow for efficient inference.

## 5. SCALING INFERENCE IN HOLOCLEAN

To repair an input dataset $D$, HoloClean grounds the DDlog generated by its compilation module and runs Gibbs sampling [37] to perform inference. Both grounding and Gibbs sampling introduce significant challenges when applied over complex probabilistic models: (i) Grounding is prone to combinatorial explosion in the presence of complex inference rules over random variables with large domains [27]; and (ii) in the presence of complex correlations, Gibbs sampling require an exponential number of iterations in the number of random variables to mix, i.e., reach a stationary distribution, and accurately estimate the marginal probabilities of query variables [36]. In this section, we introduce three optimizations to address these challenges: We first introduce two optimizations to limit the combinatorial explosion during grounding, and then propose an optimization that guarantees $O(n \log n)$ iterations for Gibbs sampling to mix, where $n$ is the number of random variables.

### 5.1 Scalable Grounding in HoloClean

The size of the factor graph generated by HoloClean may suffer from combinatorial explosion due to complex DDlog rules that correlate multiple random variables with large domains. Specifically for HoloClean, this problem rises due to the presence of factors that encode dependencies due to denial constraints (see Section 4). We consider the DDlog rule in Example 4 to demonstrate this problem:

**Example 5.** *Consider an input instance, such that all random variables associated with "Zip" take values from a domain Z, all random variables for "State" take values from a domain S, and there are T tuples in D. Given the constraints $z1 = z2$ and $s1 \neq s2$, we have that the total number of groundings just for all tuples is $O(|T|^2 \cdot |Z| \cdot |S|^2)$. The combinatorial explosion is apparent for large values of either $|S|$ or $|T|$.*

As Example 5 demonstrates, there are two aspects that affect HoloClean's scalability: (i) random variables with large domains (e.g., $|S|$ in our example), and (ii) factors that express correlations across all pairs of tuples in $D$ (e.g., $|T|$ in our example). We introduce two optimizations: (i) one for pruning the domain of random variables in HoloClean's model by leveraging co-occurrence statistics over the cell values in $D$, and (ii) one for pruning the pairs of tuples over which denial constraints are evaluated.

### 5.1.1 Pruning the Domain of Random Variables

Each cell $c$ in $D$ corresponds to a random variable $T_c$. As described in Section 2, these random variables are separated in evidence variables whose value is fixed—these correspond to clean cells in $D_c$—and query variables whose value needs to be inferred—these correspond to noisy cells in $D_n$. While the domain of evidence variables is fixed, we need to determine the domain of query random variables. Without external domain knowledge, data repairing algorithms usually allow a cell to obtain any value from the active domain of their corresponding attribute, namely, the values that have appeared in the corresponding attribute of that cell [7, 12].

In HoloClean, we use a different strategy to determine the domain of random variables $T_c$: Consider a cell $c \in D_n$ and let $t$ denote its tuple. We consider the values that other cells in tuple $t$ take. Let $c'$ be a cell in $t$ different than $c$, $v_{c'}$ its value, and $A_{c'}$ its corresponding attribute. We consider candidate repairs for $c$ to be all values in the domain of $c$'s attribute, denoted $A_c$, that co-occur with value $v_{c'}$. To limit the set of candidate repairs we only consider values that co-occur with a certain probability that exceeds a pre-defined threshold $\tau$. Following a Bayesian analysis we have: Given a threshold $\tau$ and values $v$ for $A_c$ and $v_{c'}$ for $A_{c'}$, we require that the two values co-occur if $Pr[v|v_{c'}] \geq \tau$. We define this as:

$$Pr[v|v_{c'}] = \frac{\#(v, v_{c'}) \text{ appear together in } D}{\#v_{c'} \text{ appears in } D}$$

The overall candidate discovery algorithm used in HoloClean is shown in Algorithm 2.

---
**Algorithm 2:** Domain Pruning of Random Variables

**Input**: Set of Noisy Data Cells $D_n$, Dataset $D$, Threshold $\tau$
**Output**: Repair Candidates for Each Cell in $D_n$
**for** *each cell c in $D_n$* **do**
  /* Initialize repair candidates for cell $c$ */
  $R_c \leftarrow \emptyset$;
  **for** *each cell c in $D_n$* **do**
    $A_c \leftarrow$ the attribute of cell $c$;
    **for** *each cell $c' \neq c$ in $c$'s tuple* **do**
      $U_{A_c} \leftarrow$ the domain of attribute $A_c$;
      $v_{c'} \leftarrow$ the value of cell $c'$;
      **for** *each value $v \in U_{A_c}$* **do**
        **if** $Pr[v|v_{c'}] \geq \tau$ **then**
          $R_c \leftarrow R_c \cup \{v\}$;

return repair candidates $R_c$ for each $c \in D_n$;

---

Varying threshold $\tau$ in Algorithm 2 allows us to prune the domain of random variables and systematically trade-off the scalability of HoloClean and the quality of repairs obtained by it. In our experiments (see Section 6) we find that varying the value of $\tau$ not only affects the runtime of HoloClean but also introduces a trade-off between the precision and recall of repairs output by HoloClean. We also find our pruning strategy to be necessary for HoloClean to scale to large data cleaning instances—the largest dataset we considered contains 2.7 million tuples.

### 5.1.2 Tuple Partitioning Before Grounding

Grounding the DDlog rules generated by Algorithm 1 requires iterating over all pairs of tuples in $D$ and evaluating if the body of each DDlog rule is satisfied for the random variables corresponding to their cells. However, in practice, there are many tuples that will never participate in a constraint violation (e.g., the domains of their cells may never overlap). To avoid evaluating DDlog rules for such tuples, we introduce a scheme that partitions an input dataset $D$ in groups of tuples such that tuples in the same group have a high probability of participating in a constraint violation. DDlog rules are evaluated only over these groups, thus, limiting the quadratic complexity of grounding the rules generated by Algorithm 1.

To generate these groups we leverage *conflict hypergraphs* [26] which encode constraint violations in the original dataset $D$. Nodes in conflict hypergraph $H$ correspond to cells that participate in detected violations and hyperedges link together cells involved in the same violation. Hyperedges are also annotated with the constraint that generated the violation.

For each constraint $\sigma \in \Sigma$ we consider the subgraph of $H$ containing only hyperedges for violations of $\sigma$. Let $H_\sigma$ be the induced subgraph. We let each connected component in $H_\sigma$ define a group of tuples over which the factor for constraint $\sigma$ will be materialized. The overall process is described in Algorithm 3.

---
**Algorithm 3:** Generating Tuple Groups

**Input**: Dataset $D$, Constraints $\Sigma$, Conflict Hypergraph $H$
**Output**: Groups of Tuples
/* Initialize set of tuple groups */
$G \leftarrow \emptyset$;
**for** *each constraint $\sigma$ in $\Sigma$* **do**
  $H_\sigma \leftarrow$ subgraph of $H$ with violations of $\sigma$;
  **for** *each connected component $cc$ in $H_\sigma$* **do**
    $G \leftarrow G \cup \{(\sigma, \text{tuples from } D \text{ present in } cc)\}$;

return set of tuple groups $G$;

---

We us the output of Algorithm 3 to restrict the groundings of rules generated by Algorithm 1 only over the tuples that belong in the same connected component with respect to each denial constraint $\sigma \in \Sigma$. Our partitioning scheme limits the number of factors generated due to denial constraints to $O(\sum_{g \in G} |g|^2)$ as opposed to $O(|\Sigma||D|^2)$. In the worst case the two quantities can be the same. In our experiments, we observe that, when random variables have large domains, i.e., a small value of $\tau$ is used in Algorithm 2, our partitioning optimization leads to more scalable models—we observe speed-ups up to $2\times$—that output accurate repairs; compared to inference without partitioning, we find an F1-score decrease of 6% in the worst case and less than 0.5% on average.

## 5.2 Rapid Mixing of Gibbs Sampling

Gibbs sampling requires that we iterate over the random variables in the factor graph and, at every step, sample a single variable from its conditional distribution (i.e., keep all other variables fixed).

Unfortunately, for complex factor graphs, Gibbs sampling requires an exponential number of iterations in the number of variables to mix, i.e., reach a stationary distribution [36]. However, if a factor graph has only independent random variables then Gibbs sampling requires $O(n \log n)$ steps to mix [21, 36].

Motivated by this result, we introduce an optimization that relaxes the DDlog rules generated by Algorithm 1 to obtain a model with independent random variables. Instead of enforcing denial constraints for any assignment of the random variables corresponding to noisy cells in $D$, we use the available integrity constraints to generate features that provide evidence on random variable assignments that lead to constraint violations. To this end, we introduce an approximation of our original probabilistic model that builds upon two assumptions: (i) erroneous cells in $D$ are fewer than correct cells, i.e., for each error there is sufficient information in $D$ to repair it, and (ii) each integrity constraint violation can be fixed by updating a single cell in the participating tuples.

We relax the DDlog rules generated by Algorithm 1 using the following procedure: For each rule in the output of Algorithm 1, Iterate over each Value?() predicate and for each such predicate generate a new DDlog whose head contains only that while all remaining Value?() predicates are converted to InitValue() predicates in the body of the rule. Also the weights of the original rules are relaxed to learnable parameters of the new model.

The above procedure decomposes each initial DDlog rule into a series of new rules whose head contains a single random variable. We use an example to demonstrate the output of this procedure.

**Example 6.** *We revisit Example 4. Our approximation procedure decomposes the initial DDlog rule into the following rules:*

!Value?(t1, Zip, z1) : −InitValue(t2, Zip, z2),

InitValue(t1, State, s1), InitValue(t2, State, s2)),

Tuple(t1), Tuple(t2), $[t1! = t2, z1 = z2, s1 \neq s2]$ $weight = w$

*and*

!Value?(t1, State, s1) : −InitValue(t1, Zip, z1),

InitValue(t2, Zip, z2), InitValue(t2, State, s2)),

Tuple(t1), Tuple(t2), $[t1! = t2, z1 = z2, s1 \neq s2]$ $weight = w$

*where in contrast to the fixed weight of the original rule, w for the two rules above is a weight to be estimated during learning.*

In contrast to HoloClean's initial probabilistic model, the approximate model do not penalize arbitrary violations of constraints in $\Sigma$ that may occur if the values of multiple cells are updated at the same time. Despite the latter, our relaxed model comes with two desired properties: (i) the factor graph generated by relaxing the original DDlog rules contains only independent random variables, hence, Gibbs sampling is guaranteed to mix in $o(n \log n)$ steps, and (ii) since random variables are independent learning the parameters of Equation 1 corresponds to a convex optimization problem. In Section 6.3, we demonstrate that this model not only leads to more scalable data repairing methods but achieves the same quality repairs as the non-relaxed model.

# 6. EXPERIMENTS

We compare HoloClean against state-of-the-art data cleaning methods on a variety of synthetic and real-world datasets. We show that HoloClean yields an average F1 improvement of more than 2×. The main points we seek to validate are: (i) how accurately can HoloClean repair real-world datasets containing a variety of errors, (ii) what is the impact of different signals on data repairing, and (iii) what is the impact of our pruning methods on the scalability and accuracy of HoloClean.

**Table 2: Parameters of the data used for evaluation. Noisy cells do not necessarily correspond to erroneous cells.**

| Parameter | Hospital | Flights | Food | Physicians |
|---|---|---|---|---|
| Tuples | 1,000 | 2,377 | 339,908 | 2,071,849 |
| Attributes | 19 | 6 | 17 | 18 |
| Violations | 6,604 | 84,413 | 39,322 | 5,427,322 |
| Noisy Cells | 6,140 | 11,180 | 41,254 | 174,557 |
| ICs | 9 DCs | 4 DCs | 7 DCs | 9 DCs |

## 6.1 Experimental Setup

We describe the datasets, metrics, and experimental settings used to validate HoloClean against competing data repairing methods.

*Datasets.* We use four real data sets: one commonly used hospital data set for evaluating data repairing, one with flight scheduling data, the food inspection dataset from Chicago's data catalog (Section 1), and a dataset with information on medical professionals from Medicare.gov. For all datasets we seek to repair cells that participate in violations of integrity constraints. Table 2 shows statistics for these datasets. As shown, the datasets span different sizes and exhibit various amounts of errors:

*Hospital.* This is a typical benchmark dataset used in the data cleaning literature [12, 14]. Errors amount to ∼ 5% of the total data. Ground truth information is available for all cells. This is an easy benchmark with significant duplication across cells. *We use this dataset to evaluate how effective HoloClean is at leveraging the presence of duplicate information—across cells that may not even participate in violations—during cleaning.*

*Flights.* This dataset [30] contains information on the departure and arrival time of flights as reported by different data sources on the web. We use four denial constraints that ensure a unique scheduled and actual departure and arrival time for each flight. Errors arise due to conflicts across data sources. Ground truth information is available for all cells. The majority of cells in Flights are noisy. Flights contains information on which source provides each tuple. *We use this dataset to examine how robust HoloClean is in the presence of large numbers of erroneous cells, and to evaluate if HoloClean can successfully exploit conflicts across data sources to identify correct data repairs.*

*Food.* This is the dataset from Example 1. It contains information on food establishments in Chicago and was obtained by transcribing forms filled out by city inspectors. Errors correspond to conflicting zip codes for the same establishment, conflicting inspection results for the same establishment on the same day, conflicting facility types for the same establishment and many more. These errors are captured by seven denial constraints. In this dataset the majority of errors are introduced in non-systematic ways. The dataset also contains many duplicates as records span different years. *We use this dataset to evaluate HoloClean against real-life data with duplicate information and non-systematic errors.*

*Physicians.* This is the Physician Compare National dataset published in Medicare.gov. [2] It contains information on medical professionals and the primary organization they are associated with. We used nine denial constraints to identify errors in the dataset. The majority of errors correspond to systematic errors. For example, the location field for medical organizations is misspelled, thus, introducing systematic errors across entries of different professionals. For instance, "Sacramento, CA" is reported as "Scaramento, CA" in 321 distinct entries. Other systematic errors include zip code to state inconsistencies. *We use this dataset to evaluate HoloClean against datasets with systematic errors across multiple entries.*

---

[2] https://data.medicare.gov/data/physician-compare

*Competing Methods.* For the experiments in this section, denial constraints in HoloClean are relaxed to features that encode priors over independent random variables (see Section 5.2). No partitioning is used. We evaluate HoloClean against three competing data repairing approaches:

- **Holistic** [12]: This holistic repairing method leverages denial constraints to repair erroneous records. Holistic is shown to outperform other methods based on logical constraints, thus, we choose to compare HoloClean against this method alone.
- **KATARA** [13]: This is a knowledge base (KB) powered data cleaning system that, given a dataset and a KB, interprets table semantics to align it with the KB, identifies correct and incorrect data, and generates repairs for incorrect data. We use an external dataset containing a list of States, Zip Codes, and Location information as external information.[3]
- **SCARE** [39]: This is a state-of-the-art data cleaning method that relies on machine learning and likelihood methods to clean dirty databases by value modification. This approach does not make use of integrity or matching constraints.

*Features, Error Detection, and External Signals.* The probabilistic models generated by HoloClean capture all features described in Section 4. Source-related features are only available for Flights. To detect erroneous cells in HoloClean, we used the same mechanism as Holistic [12]. Finally, for micro-benchmarking purposes we use the dictionary used for KATARA on Hospital, Food, and Physicians. Unless explicitly specified HoloClean does not make use of this external information.

*Obtaining Groundtruth Data.* Food and Physicians come with no ground truth. To evaluate data repairing on these datasets, we manually labeled 2,000 and 2,500 cells, respectively: We focused on tuples identified as erroneous by the error detection mechanisms of Holistic, KATARA, and SCARE. From this set of cells we randomly labeled 2,000 cells for Food and 2,500 cells for Physician. Not all cells were indeed noisy. The labeled data obtained via this process lead to unbiased estimates for the precision of each method. However, recall measurements might be biased.

*Evaluation Methodology.* To measure the quality of the repairs performed by different methods we leverage the presence of labeled data and use:

- **Precision (Prec.)**: the fraction of correct repairs, i.e., repairs that match the ground truth, over the total number of repairs performed. For Food and Physicians the precision obtained by evaluating on the labeled examples is unbiased.
- **Recall (Rec.)**: correct repairs over the total number of errors.
- **F1-score (F1)**: the harmonic mean of precision and recall computed as $2 \times (P \times R)/(P + R)$.

For each method we also measure the *wall-clock runtime*. For HoloClean this is: (i) the time for detecting violations, (ii) the time for compiling the input noisy database to a probabilistic model, and (iii) the time needed to run learning and inference, i.e., perform data repairing. Finally, we vary the threshold $\tau$ of our pruning optimization for determining the domain of cell-related random variables (see Algorithm 2) in $\{0.3, 0.5, 0.7, 0.9\}$.

[3] This dataset was downloaded from federalgovernmentzipcodes.us

Table 3: Precision, Recall and F1-score for different datasets. Bold indicates the best performing method. For each dataset, the threshold used for pruning the domain of random variables is reported in parenthesis.

| Dataset ($\tau$) | Metric | HoloClean | Holistic | KATARA | SCARE |
|---|---|---|---|---|---|
| Hospital (0.5) | Prec. | **1.0** | 0.517 | 0.983 | 0.667 |
|  | Rec. | **0.713** | 0.376 | 0.235 | 0.534 |
|  | F1 | **0.832** | 0.435 | 0.379 | 0.593 |
| Flights (0.3) | Prec. | **0.887** | 0.0 | n/a | 0.569 |
|  | Rec. | **0.669** | 0.0 | n/a | 0.057 |
|  | F1 | **0.763** | 0.0[*] | n/a | 0.104 |
| Food (0.5) | Prec. | **0.769** | 0.142 | 1.0 | 0.0 |
|  | Rec. | **0.798** | 0.679 | 0.310 | 0.0 |
|  | F1 | **0.783** | 0.235 | 0.473 | 0.0[+] |
| Physicians (0.7) | Prec. | **0.927** | 0.521 | 0.0 | 0.0 |
|  | Rec. | **0.878** | 0.504 | 0.0 | 0.0 |
|  | F1 | **0.897** | 0.512 | 0.0[#] | 0.0[+] |

[*] Holistic did not perform any correct repairs.
[+] SCARE did not terminate after three days.
[#] KATARA performs no repairs due to format mismatch for zip code.

*Implementation Details.* HoloClean's compiler is implemented in Python while the inference routines are executed in DeepDive v0.9 using Postgres 9.6 for backend storage. Holistic is implemented in Java and uses the Gurobi Optimizer 7.0 as its external QP tool [4]. KATARA and Scare are also implemented in Java. All experiments were executed on a machine with four CPUs (each CPU is a 12-core 2.40 GHz Xeon E5-4657L), 1TB RAM, running Ubuntu 12.04. While all methods run in memory, their footprint is significantly smaller than the available resources.

## 6.2 Experimental Results

We compare HoloClean with competing data repairing approaches on the quality of the proposed repairs. We find that in all cases HoloClean outperforms all state-of-the-art data repairing methods and yields an average F1-score improvement of more than $2\times$.

### 6.2.1 Identifying Correct Data Repairs

We report the precision, recall, and F1-score obtained by HoloClean and competing approaches. The results are shown in Table 3. For each dataset, we report the threshold $\tau$ used for pruning the domain of random variables (see Algorithm 2). The effect of $\tau$ on the performance of HoloClean is studied in Section 6.3.1. As shown in Table 3 HoloClean outperforms other data repairing methods significantly with relative F1-score improvements of more than $40\%$ in all cases. This verifies our hypothesis that unifying multiple signals leads to more accurate automatic data cleaning techniques.

We focus on HoloClean's performance for the different datasets. For Hospital, HoloClean leverages the low number of errors and the presence of duplicate information to correctly repair the majority of errors, achieving a precision of 100% and a recall of 71.3%. HoloClean also achieves high precision for Flights (88.8%), as it uses the information on which source provided which tuple to estimate the reliability of different sources [35] and leverages that to propose repairs. Nonetheless, we see that recall is limited (66.9%) since most of the cells contains errors. Finally, for Food and Physician HoloClean obtains F1-scores of 0.783 and 0.897, respectively.

We now turn our attention to competing methods. We start with Holistic, which relies only on logical constraints and performs repairs to individual cells iteratively until no constraints are violated. We find that this approach yields repairs of fair quality (around 50% F1) for datasets with a large number of duplicate information (e.g.,

[4] https://www.gurobi.com/

**Table 4: Runtime analysis of different data cleaning methods. A dash indicates that the system failed to terminate after a three day runtime threshold.**

| Dataset | HoloClean | Holistic | KATARA | SCARE |
|---|---|---|---|---|
| Hospital | 147.97 sec | 5.67 sec | 2.01 sec | 24.67 sec |
| Flights | 70.6 sec | 80.4 sec | n/a | 13.97 sec |
| Food | 32.8 min | 7.6 min | 1.7 min | - |
| Physicians | 6.5 hours | 2.03 hours | 15.5 min | - |

Hospital) or a large number of systematic errors (e.g., Physicians). On the other hand, when datasets contain mostly noisy cells (as in Flights) or errors that follow random patterns (as in Food) using logical constraints and minimality as the only principle for data repairing yields very poor results—the precision of performed repairs is $0.0$ for Flights and $0.14$ for Foods.

In contrast, KATARA obtains repairs of very high precision but limited recall. This is expected as the coverage of external knowledge bases can be limited. Finally, SCARE performs reasonably well in datasets such as Hospital, where a large number of duplicate records is available and qualitative statistics can help repair errors. Similar to HoloClean it is able to leverage existing correct tuples to perform repairs. However, for Flights, where the number of duplicates is limited, SCARE has limited recall. Also SCARE failed to terminate after running for three days on Food and Physicians.

*Takeaways.* HoloClean's holistic approach obtains data repairs that are significantly more accurate—we find an F1-score improvement of more than $2\times$ on average—than existing state-of-the-art approaches that consider isolated signals for data repairing.

### 6.2.2 Runtime Overview

We measure the total wall-clock runtime of each data repairing method for all datasets. The results are shown in Table 4. Reported runtimes correspond to end-to-end execution with data pre-processing and loading. For Holistic, pre-processing corresponds to loading input data from raw files and running violation detection. SCARE operates directly on the input database, while KATARA loads data in memory and performs matching and repairing.

As shown HoloClean can scale to large real-world data repairing scenarios. Focusing on small datasets, i.e., Hospital and Flights, we find that the total execution time of HoloClean is within one order of magnitude of Holistic's runtime but still only a few minutes in total. For Food, HoloClean exhibits a significantly higher runtime but for Physicians both systems are within the same magnitude. KATARA is significantly faster as it only performs matching operations. Finally, while SCARE is very fast for the small datasets, it fails to terminate for the larger ones. While HoloClean's runtime is in general higher than that of competing methods, the accuracy improvements obtained by using HoloClean justify the overhead.

## 6.3 Micro-benchmark Results

We evaluate the tradeoff between the runtime of HoloClean and the quality of repairs obtained by it due to the optimizations in Section 5. We also evaluate the quality of data repairs performed by HoloClean when external dictionaries are incorporated.

### 6.3.1 Tradeoffs Between Scalability and Quality

We evaluate the runtime-quality tradeoffs for: (i) pruning the domain of random variables, which restricts the domain of the random variables in HoloClean's model, (ii) partitioning, and (iii) relaxing the denial constraints to features that encode priors. Domain prun-ing can be applied together with the other two optimizations, thus, is applicable to all variations of HoloClean listed next:

- **DC Factors**: Denial constraints are encoded as factors (see Section 4). No other optimization is used.
- **DC Factors + partitioning**: Same as the above variation with partitioning (see Section 5.1.2).
- **DC Feats**: Denial constraints are used to extract features that encode priors over independent random variables (see Section 5.2). This version of HoloClean was used for the experiments in Section 6.2.
- **DC Feats + DC Factors**: We use denial constraints to extract features that consider only the initial values of the values in $D$ and also add factors that enforce denial constraints for any assignment of the cell random variables.
- **DC Feats + DC Factors + partitioning**: Same as the above variation with partitioning.

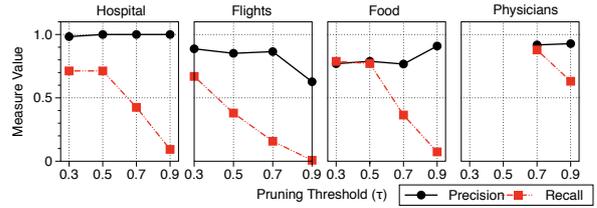

**Figure 3: Effect of pruning on Precision and Recall. Missing values correspond to time-outs with a threshold of one day.**

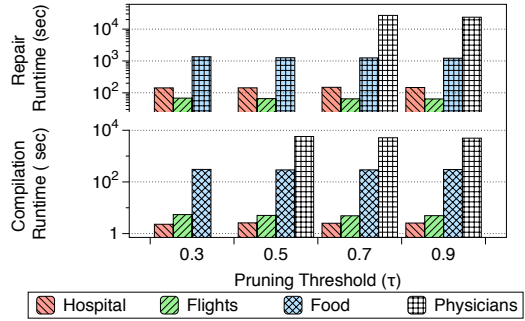

**Figure 4: Effect of pruning on Compilation and Repairing runtimes. Runtimes are reported in log-scale. Missing values correspond to time-outs with a threshold of one day.**

**The Effect of Domain Pruning.** First, we consider the DC Feats variation of HoloClean and vary threshold $\tau$. We examine how the precision and recall of HoloClean's repairs change as we limit the number of possible repairs considered. The results are shown in Figure 3. As expected, increasing the pruning threshold $\tau$ in Algorithm 2 introduces a *tradeoff between the precision and recall* achieved by HoloClean. Lower values of threshold $\tau$ provide HoloClean with an increased search space of possible repairs, thus, allowing the recall of HoloClean to be higher.

As we increase threshold $\tau$ the recall of HoloClean's output drops significantly. For example, in Food increasing the pruning threshold from $0.5$ to $0.7$ has a dramatic effect on recall, which drops from $0.77$ to $0.36$. On the other hand, we see that precision increases. One exception is the Flights dataset where a large number of the pruning threshold has a negative impact on precision. This result is expected since Flights contains a small number of duplicates: Setting $\tau = 0.9$ requires that a candidate value for a cell has high co-occurrence probability with the values that other cells in the same tuple obtain. In the absence of noisy duplicates, severe pruning can lead to a set of candidate assignments that may not contain the truly correct value for an erroneous cell.

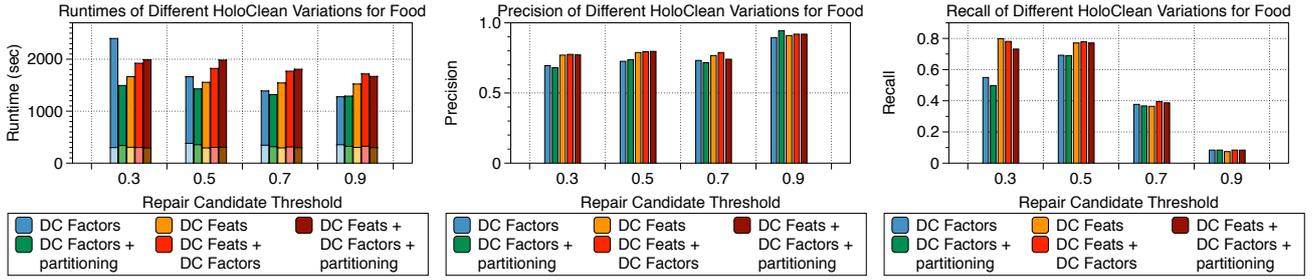

Figure 5: Runtime, precision, and recall for all variations of HoloClean on Food. For runtime, the lower part of the stacked bars corresponds to compilation time while the upper part to the runtime required for learning and inference.

The effect of the pruning threshold $\tau$ on the runtime of HoloClean is shown in Figure 4. Violation detection is not affected by this threshold, thus, we focus on the compile and repair phase. The corresponding runtimes are in log-scale due to account for dataset differences. As shown the effect of $\tau$ is not that significant on the runtime of HoloClean. Compilation runtime is similar as $\tau$ varies. However, the time required for repairing decreases as threshold $\tau$ increases and this allows HoloClean to perform accurate repairs to large datasets such as Physicians (containing $37M$ cells).

*Takeaways.* Our domain pruning strategy plays a key role in achieving highly accurate repairs and allows HoloClean to scale to large datasets with millions of rows.

**Runtime versus Quality Tradeoff.** We now evaluate the runtime, precision, and recall for all variations of HoloClean listed above. Figure 5 reports the results for Food. The same findings hold for all datasets. We make the following observations:

**(1) Runtime:** When random variables are allowed to have large domains (i.e., for small values of $\tau$) using partitioning or relaxing denial constraints to features (DC Feats) lead to runtime improvements of up to 2x. On the other hand, when the domain of random variables is heavily pruned, all variants of HoloClean exhibit comparable runtimes. This is expected as the underlying inference engine relies on database optimizations, such as indexing, to perform grounding. Not surprisingly, we see that encoding denial constraints as factors (DC Factors) instead of features (DC Feats) exhibits a better runtime. This is because the model for DC Factors contains a fewer number of factors—recall that relaxing denial constraints to features introduces a separate factor for each attribute predicate in a constraint. While one would expect partitioning to have a significant impact on the time required to perform grounding, we find that limiting the number of possible repairs per records is more effective at speeding-up grounding. The reason is that modern probabilistic inference engines leverage database optimizations such as indexing during grounding.

**(2) Quality of Repairs:** We observe that for all methods pruning the domain of random variables leads to an increase in the precision and a decrease in the recall of repairs obtained for the different variants of HoloClean. An interesting observation is that relaxing denial constraints and leveraging the initial violations to obtain evidence on the assignment of noisy cells (e..g, when DC Feats is used), allows HoloClean to obtain higher quality repairs. We conjecture that this is due to two reasons: (i) the fact that the input noisy datasets are statistically close to their true clean versions, i.e., the noise is limited, and (ii) when the domain of random variables is misspecified (e.g., too large) using a complex model that enforces denial constraints leads to harder, ill-posed inference problems. The latter is supported by the fact that combining denial constraint factors with denial constraint features improves the quality of repairs. Conducting a formal theoretical study of the aforementioned scenarios is an exciting future research direction.

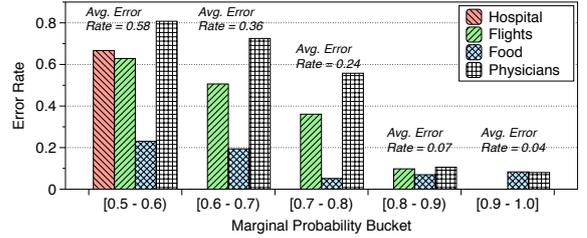

Figure 6: The error-rate of HoloClean repairs for different probability buckets for different datasets.

*Takeaways.* Relaxing denial constraints leads to more scalable models and models that obtain higher quality repairs when the domains of random variables are misspecified. A theoretical study of when encoding denial constraints as features is sufficient to obtain high quality repairs is an exciting future direction of research.

### 6.3.2 External Dictionaries in HoloClean

Finally, we evaluate the performance of HoloClean when incorporating external dictionaries and use matching dependencies. We use the same dictionary used for KATARA. The dictionary contains a list of Zip codes, cities, and states in the US. We found that using external dictionaries can improve the quality of repairs obtained by HoloClean but the benefits are limited: for all datasets we observed F1-score improvements of less than 1%. This limited gain is not a limitation of HoloClean, which can natively support external data, but due to the limited coverage of the external data used.

### 6.3.3 Qualitative Analysis on Real-Data

We perform a qualitative analysis to highlight how the marginal probabilities output by HoloClean allows users to reason about the validity of different repairs obtained by HoloClean, thus, obviating the need for exploration strategies based on active learning. We conduct the following experiment: we consider the repairs suggested by HoloClean and measure the error-rate (i.e., the rate of correct versus total repairs) for repairs in different buckets of marginal probabilities. We use the same setup as in Section 6.2.1.

As shown in Figure 6, the error-rate rate decreases as the marginal probabilities increase. For example, repairs whose marginals belong in the $[0.5 - 0.6)$ probability bucket exhibit an average error rate of $0.58$ across all datasets, while marginals in $[0.7-0.8)$ bucket have an average error rate of $0.24$. These marginal probabilities can be used to control the quality of repairs by HoloClean.

## 7. CONCLUSIONS

We introduced HoloClean, a data cleaning system that relies on statistical learning and inference to unify a range of data repairing methods under a common formal framework. We introduced several optimization to scale inference for data repairing, and studied the tradeoffs between the quality of repairs and runtime of HoloClean for those optimizations. We empirically showed that HoloClean obtains repairs that are significantly more accurate that state-of-the-art data cleaning methods.

Our study introduces several exciting research directions. Understanding when integrity constraints need to be enforced versus when it is sufficient to encode them as features has the potential to generate a new family of data repairing tools that not only scale to large instances but also come with rigorous theoretical guarantees. Additionally, data cleaning is limited by the error detection methods used before. Recently, the paradigm of data programming [34] has been introduced as a means to allow users to programmatically encode domain knowledge in inference tasks. Exploring how data programming and data cleaning can be unified under a common probabilistic framework to perform better detection and repairing is a promising future direction.

## 8. ACKNOWLEDGEMENTS

The authors would like to thank the members of the Hazy Group for their feedback and help. We would like to thank Intel, Toshiba, the Moore Foundation, and Thomson Reuters for their support, along with NSERC through a Discovery Grant and DARPA through MEMEX (FA8750-14-2-0240), SIMPLEX (N66001-15-C-4043), and XDATA (FA8750-12-2-0335) programs, and the Office of Naval Research (N000141210041 and N000141310129). Any opinions, findings, and conclusions or recommendations expressed in this material are those of the authors and do not necessarily reflect the views of DARPA, ONR, or the U.S. government.## 9. REFERENCES